\documentclass[12pt]{article}

\usepackage{color}

\usepackage{amsmath}

\usepackage{mathtools}
\usepackage{amsmath}
\usepackage{graphicx}
\usepackage{float}
\usepackage{amsmath}
\usepackage{amsthm}
\usepackage{amssymb}
\usepackage{amscd}
\usepackage{amsfonts}
\usepackage{makeidx}
\usepackage{enumerate}      
\usepackage{IEEEtrantools}
\usepackage{mathrsfs}

\usepackage{amsmath,empheq}

\usepackage [all]{xy}
\usepackage{cases}

\usepackage{amsmath}

\usepackage{cancel}

\usepackage{mathrsfs}

\newtheorem{teo}{Theorem}[section]
    \newtheorem{lem}[teo]{Lemma}
    \newtheorem{prop}[teo]{Proposition}

    \newtheorem{obs}[teo]{Remark}

    \newtheorem*{dem}{\textsc{Proof}}
    \newcommand{\bdem}{\begin{dem}}
    \newcommand{\edem}{\end{dem}}
     \newcommand{\be}{\begin{equation}}
    \newcommand{\ee}{\end{equation}}
     \newcommand{\ba}{\begin{array}}
    \newcommand{\ea}{\end{array}}
\newcommand{\beqn}{\begin{eqnarray}}
    \newcommand{\eeqn}{\end{eqnarray}}
    \newcommand{\bl}{\begin{lem}}
    \newcommand{\el}{\end{lem}}
    \newcommand{\bp}{\begin{prop}}
    \newcommand{\ep}{\end{prop}}
\newcommand{\ds}{\displaystyle}
     
     \newcommand{\al}{\alpha}

    \newcommand{\R}{\mathbb{R}}
    \newcommand{\C}{\mathbb{C}}

   \providecommand{\abs}[1]{\lvert#1\rvert}

    \IEEEyessubnumber

\def\Re {{\rm Re\, }}                                       
 \def\Im {{\rm Im\,}}

\begin{document}
\title{Asymptotic forms of the Ursell edge waves on a gently sloping beach}

\author{
\large{ P.  Zhevandrov$^1$},\\
\large{A.  Merzon$^2$},\\
\large{and M.I. Romero Rodr\'iguez$^3$}\\
{\small{\it $^1$ Facultad de  Ciencias F\'\i sico-Matem\'aticas, Universidad Michoac{a}na}},\\[-2mm]
{\small  Morelia, Michoac\'{a}n, M\'{e}xico}\\[-2mm]
{\small{\it $^2$ Instituto de F\'\i sica y  Matem\'aticas, Universidad Michoac{a}na}},\\[-2mm]
{\small  Morelia, Michoac\'{a}n, M\'{e}xico}\\[-2mm]
{\small{\it $^3$ Facultad de Ciencias B\'asicas y Aplicadas, Universidad Militar Nueva Granada}},\\[-2mm]
{\small{Bogot\'a, Colombia}}.
\\{\small{\it E-mails}: pzhevand@gmail.com,}
{\small anatolimx@gmail.com,}\\[-2mm] {\small maria.romeror@unimilitar.edu.co}}
\maketitle

\begin{abstract} It is shown that the asymptotics of the Ursell trapping modes on a sloping beach of small slope angle $\alpha$ coincide with the inverse Fourier transform of standard WKB exponentials, i.e., the Maslov canonical operator,   as $\alpha\to0$.
\end{abstract}


\section{Introduction}
\setcounter{equation}{0}

The Ursell trapping modes on a beach of constant slope are solutions of the following boundary value problem for the function $\varphi(x,y,\alpha)$:  
\begin{align}\label{delV}
&\Delta \varphi-\varphi=0\quad {\rm in}~~\Omega,
\\\nonumber\\\label{lam^2}
&\ds\frac{\partial\varphi}{\partial y}+\lambda\varphi=0\quad {\rm on}~~\Gamma_{F},\qquad\ds\frac{\partial \varphi}{\partial n}=0\quad {\rm on}~~\Gamma_{B},
\end{align}
 where $x$ is the horizontal coordinate, $\infty> x>0$; the $x$-axis is orthogonal to the shoreline $x=y=0$, $y$ is the downward-pointing vertical coordinate,  $\Omega=\big\lbrace x\tan\alpha>y>0\big\rbrace$, $\Gamma_{F}$ is the free surface of the liquid, $\Gamma_{F}=\big\lbrace y=0,~ \infty>x>0\big\rbrace$, $\Gamma_{B}$ is the bottom, $\Gamma_{B}=\big\lbrace y=x\tan\alpha, \infty>x>0\big\rbrace$, $\alpha$ is the magnitude of the angle $\Omega$, $\lambda=\omega^2/gk$, $k$ is the wavenumber along the $z$-axis, where $z$ is the coordinate along the shoreline;  in (\ref{delV})-(\ref{lam^2}) this coordinate (as well as the time) is absent since one assumes that the velocity potential of the initial wave problem is proportional to $\varphi(x,y,\alpha)\exp(ikz-i\omega t)$. 
Also, (\ref{delV}) and (\ref{lam^2}) are obtained after the scaling $(x,y)\longrightarrow (kx,ky)$.  

\begin{figure}[htbp]
\centering
\includegraphics[scale=0.23]{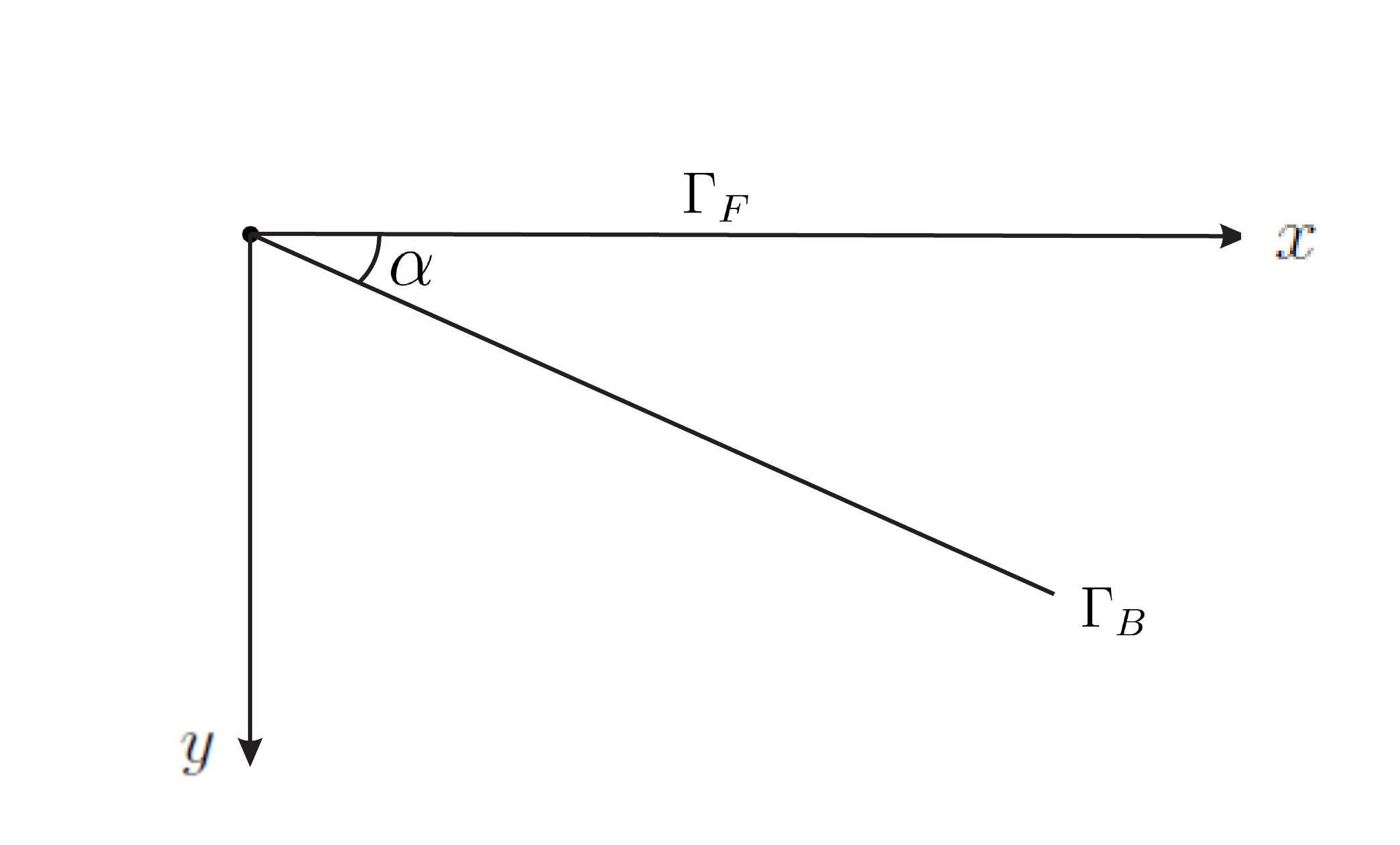}
\caption{Ursell's problem}\label{OMe}
\end{figure}

The parameter $\lambda$ in (\ref{lam^2}) is a spectral parameter of our problem and Ursell \cite{Ursell} has shown that there exist nontrivial finite energy solutions $U_n(x,y)$ of (\ref{delV})-(\ref{lam^2}) when
 \begin{equation}\label{lam_{n}^2}
\lambda= \lambda_{n}=\sin(2n+1)\alpha, \quad n=0,1,2,\dots\quad\text{such that}\quad (2n+1)\alpha<\ds\frac{\pi}{2}. 
\end{equation}     
These solutions  have the form
\begin{equation}\label{varphi_n}
\begin{array}{lll}
U_{n}(x,y)&=&\exp\Big\lbrace -\big[x\cos\alpha+y\sin\alpha\big]\Big\rbrace\\\\&+&\ds\sum_{m=1}^{n} A_{mn} \Bigg\lbrace\exp\Big\lbrace -\Big[x\cos(2m-1)\alpha-y\sin(2m-1)\alpha\Big]\Big\rbrace \\\\
&+& \exp\Big\lbrace -\Big[x\cos(2m+1)\alpha+y\sin(2m+1)\alpha\Big]\Big\rbrace\Bigg\rbrace.
\end{array}
\end{equation}
\begin{equation}\label{A_mn}
A_{mn}=(-1)^{m}\prod_{r=1}^{m}\Big[\tan(n-r+1)\alpha/\tan(n+r)\alpha\Big],\quad n=0,1,2,\cdots,\quad 1\leq m\leq n.
\end{equation}
In \cite{Mer95,KMZ} it was shown that apart from the functions $U_{n}$ there are no finite energy solutions to problem (\ref{delV})-(\ref{lam^2}).\\ 
Clearly, for small values of $\alpha$, the number of terms in the sum  (\ref{varphi_n}) tends to infinity (since  the allowed values of $n$ in (\ref{lam_{n}^2}) can be of the order of $1/\alpha$). Therefore, it is of interest to obtain asymptotic formulas for $U_{n}$ as $\alpha\to 0$. Formal ray method \cite{Keller1} and its generalizations \cite{Keller2, Zh92, SunShen}  suggest that these formulas should have the form of the WKB-solutions (quasimodes) obtained in the papers cited above. The purpose of this paper is to provide a rigorous justification of this hypothesis. 
Recall that the WKB solutions (in the momentum representation, i.e., in the form of the Maslov canonical operator, see \cite{Maslov}) have the form 
\begin{equation}\label{psi_n}
\psi_{n}(x,y)=\ds\frac{1}{2\pi}\ds\int e^{-{i\xi p}/{\alpha}+{i\Phi(p, \lambda)}/{\alpha}}\;b(p,y,\lambda)\;dp,
\end{equation}
where $\xi=\alpha x$ is the ``slow'' variable, the function $b$ is regular in $p,y,\alpha$ and the integration is carried over a suitable contour in the $p$-plane. The formal derivation of (\ref{psi_n}) is given in Appendix\;2. We note that in \cite{Zh92, SunShen} the WKB solutions were constructed with respect to the small parameter $\varepsilon=\tan\alpha$; since $\varepsilon=\alpha+O(\alpha^3)$, the leading terms for $\Phi$ and $b$ are the same for both choices of the small parameter and the leading term for $\lambda$ under the choice of $\alpha$ as a small parameter coincides with the exact formula (\ref{lam_{n}^2}).

We will show that the asymptotics of $U_{n}$ as $\alpha\to0$ coincides with the formal asymptotics (\ref{psi_n}), up to a factor, for  $n=O(1)$ as well as for $n\sim 1/\alpha$. Note that in \cite{SunShen}, in contrast to \cite{Zh92}, the WKB asymptotics was proven to be valid only for large $n$,  $n\sim 1/\alpha$. Thus here we prove that in the case of the rectilinear bottom profile the WKB solutions provide the asymptotics for the eigenvalues and the eigenfunctions. Note that the fact that formal WKB solutions provide the asymptotics of the eigenvalues is well-known (see, e.g.,  \cite[Section 13]{Maslov} and \cite[Section 7.6]{Babich}). As for the eigenfunctions,  formal asymptotics, generally speaking, do not provide their approximations \cite{Arnold, Laz}. In our case this turns out to be true.

 We note further that the results of \cite{Zh92,SunShen} are valid also for  beaches with more general bottom profile of small but nonconstant slope; this of course provides the asymptotics of the eigenvalues; we do not know whether the formal asymptotic the eigenfunctions give the asymptotics of the true eigenfunctions in this more general case.

The paper is organized as follows: in Section 2 we recall the integral representation of the Ursell modes \cite{Mer95, KMZ} and derive its asymptotics for small values of $\alpha$; this asymptotics is shown to coincide with the WKB approximation (\ref{psi_n}). In Section 3 we obtain the asymptotics of the integrals (\ref{psi_n}). In Appendices 1 and 2 we prove the integral representation of $U_n$ and recall the formal WKB method from \cite{Zh92, SunShen}.

\section{Integral representation of the Ursell modes and its asymptotics}
\setcounter{equation}{0}

We begin with the  integral representation of the Ursell modes obtained in \cite{KMZ}.
It was  was shown there (see also Appendix 1) that the functions $U_{n}$ can be expressed in the form of the Sommerfeld integrals
\begin{equation}\label{var_{x,0}}
\varphi_{n}(x,y)=\ds\frac{1}{4\pi}\ds\int\limits_{\Gamma} e^{-i\rho\sinh (w+i\theta)} P_{n}(w) dw,\quad \rho=\sqrt{x^2+y^2},\quad 0<\theta<\alpha,
\end{equation}
where
\begin{equation}\label{Ph1}
P_{n}(w)=\prod_{k=0}^{2n}\cot\Bigg(\ds\frac{\pi}{4}-\ds\frac{iw}{2}+\alpha\Big(-n+k+\ds\frac{1}{2}\Big)\Bigg),
\end{equation}
$\rho, \theta$ are the polar coordinates on the plane $(x,y)$ and the contour $\Gamma$ is the union of the line $\Gamma_{-\alpha}=\lbrace \Im w=-\alpha \rbrace$ in the positive direction and the line $\Gamma_{-\pi}=\lbrace \Im w=-\pi \rbrace$ in the negative direction (see Fig.\;2). 

\begin{figure}[htbp]
\centering
\includegraphics[scale=0.23]{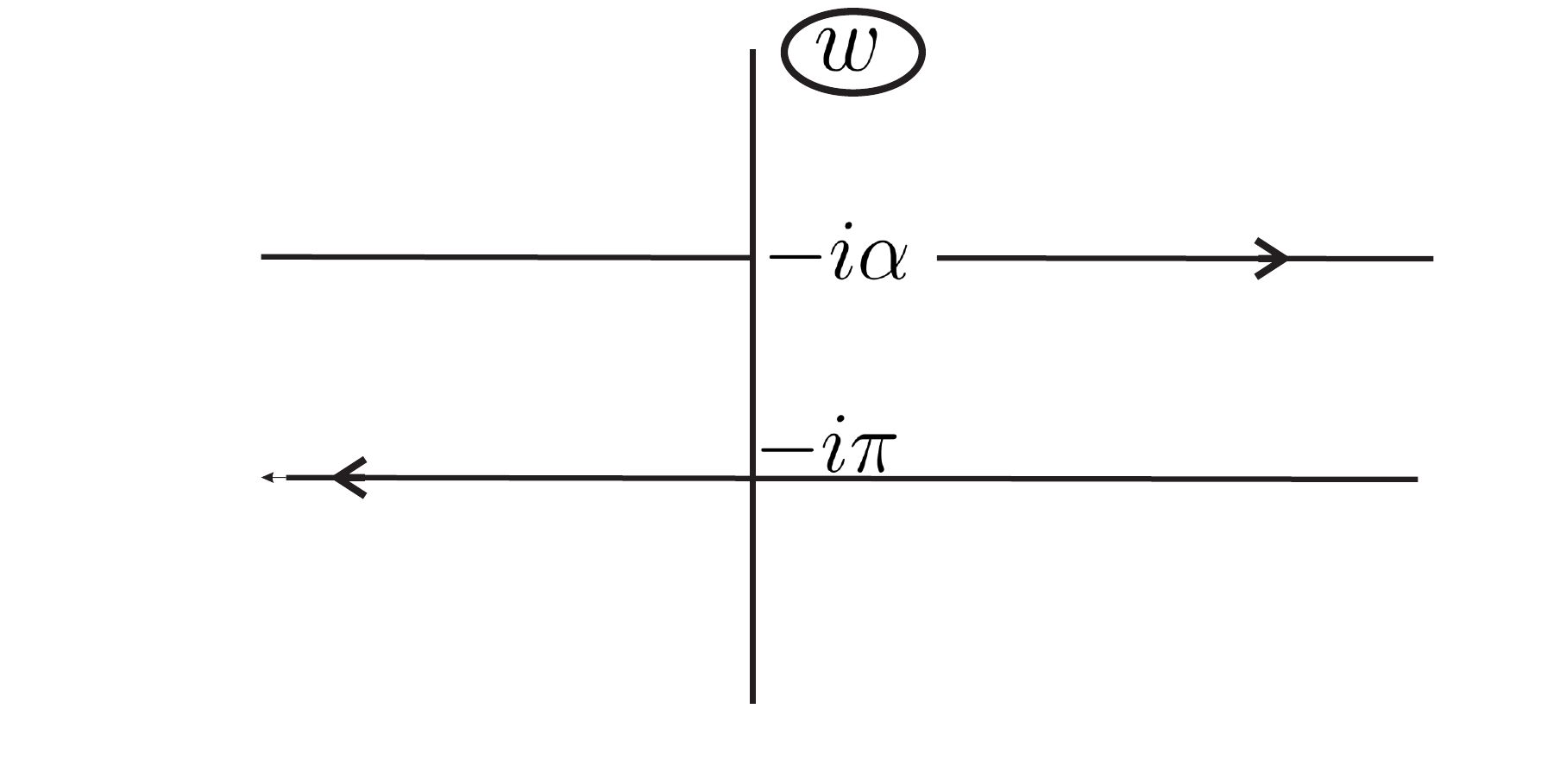}
\caption{Integration contour}\label{Gamma}
\end{figure}

More precisely, the functions $U_n$ given by (\ref{varphi_n}) satisfy 
\begin{align}\label{n>0}
& \varphi_{n}(x,y,\alpha)=\ds\frac{(-1)^{n}}{\ds\prod_{j=1}^{n}\tan^2(\alpha j)} U_{n}(x,y),\quad n=1,2,3,\cdots
\\\nonumber\\\label{nn>0}
& \varphi_0(x,y,\alpha)=U_{0}(x,y), \quad n=0.
\end{align}
Formulas (\ref{n>0}), (\ref{nn>0}) can be easily obtained  by calculating the residues of the integral in (\ref{var_{x,0}}) at zeros of the denominator (see Appendix\;1).

\begin{obs} 
In paper \cite{KMZ}  formula for $\varphi _n$  
was given in the form 
\begin{equation}\label{Var_n}
\varphi_{n}(x,y,\alpha)=\ds\frac{1}{4\pi}\ds\int\limits_{\Gamma} e^{-i\rho\sinh(w+i\theta)}\; \ds\frac{\ds\prod_{k=0}^{2n}\cos\Big(\ds\frac{\pi}{4}-i\ds\frac{w}{2}+\alpha\big(-n+k+\ds\frac{1}{2}\big)\Big)}{\ds\prod_{k=1}^{2n+1}\sin\Big(\ds\frac{\pi}{4}-i\ds\frac{w}{2}+\alpha\big(-n+k-\ds\frac{1}{2}\big)\Big)} dw
\end{equation}
It  can be easily reduced to formula (\ref{var_{x,0}}) after simple manipulations. Note that in \cite{KMZ} the corresponding formula  (see \cite[formula~(5.3)]{KMZ}) contains an erroneous factor $1/\sin\alpha$; obviously, this factor in not essential to the results of \cite{KMZ}.
\end{obs}
 Note that the  poles of the Sommerfeld kernel $P_{n}(w)$, in the domain $\big\lbrace -i\pi<\Im w<0\big\rbrace$ are
\begin{equation}\label{w_k}
w_{k}=-\ds\frac{i\pi}{2}+2i\alpha\Big(n-k-\ds\frac{1}{2}\Big),\quad k=0,1,\cdots, 2n. 
\end{equation}
The ``highest'' pole lies no higher than  $-2i\al$ and the ``lowest'' pole lies higher  than $-i\pi+2i\alpha$,  thus, on the contour $\Gamma$,  $P_n$ has neither zeros nor poles. The integral in  (\ref {var_{x,0}}) converges for 	$\rho >0$,  $0< \theta <\alpha$ since $\Im \sinh z<0$ for $\Im z \in (-\pi, 0)$ and $\theta <\al$.

Due to the fact that the poles $w_k$ can become close to the integration contour for large $n$, for technical reasons connected with the calculation of the asymptotics of $P_n$ in (\ref{var_{x,0}}) uniformly in $w$, we will assume that the numbers $n$ are such that the poles are bounded away from $\Gamma$ (see (\ref{ncond}) below).This does not mean that $n$ is bounded as $\alpha\to0$ but  this \emph{does} mean that we consider the values of $n$ such that the values of $\lambda_n$ are bounded away from the continuous spectrum $[1,\infty)$.

Obviously,
\begin{equation}\label{Fin}
\varphi_{n}(x,y)=\varphi_{n}^{(1)}(x,y)+\varphi_{n}^{(2)}(x,y),
\end{equation}
where
\begin{equation}\label{fin1}
\varphi_{n}^{(1)}(x,y):=\ds\frac{1}{4\pi}\ds\int\limits_{\underrightarrow{\Gamma_{-\alpha}}} e^{-i\rho\sinh (w+i\theta)}\; P_{n}(w)\;dw,
\end{equation}
 \begin{equation}\label{phn2}
\varphi_{n}^{(2)}(x,y):=\ds\frac{1}{4\pi}\ds\int\limits_{\underleftarrow{\Gamma_{-\pi}}} \; e^{-i\rho\sinh (w+i\theta)}\; P_{n}(w)\; dw, 
\end{equation}
and $P_{n}(w)$ is given by (\ref{Ph1}).\\
 We note that another integral representation of the Ursell modes was obtained in \cite[Section 7.5]{Whitham}. It is more useful for us to use the representation (\ref {var_{x,0}}) since it resembles the inverse Fourier transform and hence is closer to the WKB solutions (\ref{psi_n}).

\subsection{Asymptotics of $\varphi_{n}^{(1)}$}

In order to obtain the asymptotics of (\ref{var_{x,0}}), we will calculate the asymptotics of the function $P_{n}(w)$, more precisely, of its logarithm. The logarithm of $P_{n}$ turns out to be the Riemann sum of the corresponding integral.\\ 
Let 
\begin{equation}\label{fs}
f(s,w)=\ln\cot\Big(\ds\frac{\pi}{4}-\ds\frac{iw}{2}+s\Big),\quad s\in[-\mu_n,\mu_n],\quad \mu_n=\alpha\Big(n+\ds\frac{1}{2}\Big),\quad w\in\Gamma_{-\al};
\end{equation}
here and everyehre below the function $\ln(\cdot)$ is the principal branch of the logarithm.

In order that the singularities of the function $f(s,w)$ with respect to $w$ and for $s\in[-\mu_n,\mu_n]$ be bounded away from the contour of integration we will assume that the numbers $n$ satisfy the condition
\begin{equation}\label{ncond}
(2n+1)\alpha\leq\frac\pi2-\delta
\end{equation}
with an arbitrary positive $\delta$ not depending on $\alpha$. In this case, as it is easy to see, the poles $w_s$ of $\cot(\pi/4-iw/2+s)$ satisfy
\begin{equation}\label{p}
\Im w_s\in [-\pi+\delta,-\delta],\qquad s\in[-\mu_n,\mu_n];
\end{equation}
These poles are bounded away from $\Gamma_{-\alpha}$ and $\Gamma_{-\pi}$.

Note that $\lambda_{n}=\sin 2\mu_n$. Clearly, $\mu_n$ depends on $n$ but we will not emphasize this explicitly for brevity. Obviously, $f(s,w) $ is in fact a function of one argument $\pi/4-iw/2+s$, but in what follows we will need the derivatives of $f$ with respect to both $s$ and $w$; therefore we will use the notation of (\ref{fs}).

By (\ref{Ph1}) we have 
\begin{equation}\label{Pn}
\ln P_{n}(w)=\ds\sum_{k=0}^{2n} f(s_{k+1},w),\quad s_{k}=\alpha\Big(k-n-\ds\frac{1}{2}\Big).
\end{equation}
We will show that the asymptotics of the sum in (\ref{Pn}) can be expressed in terms of the integral $\ds\int\limits_{-\mu}^{\mu} f(s,w) \;ds$. This is the central technical tool which allows us to reduce the kernel $P_n$ to a WKB-type function. Indeed, 
\begin{equation}\label{is}
\ds\int\limits_{-\mu}^{\mu} f(s,w) \;ds=\sum_{k=0}^{2n} \ds\int\limits_{s_{k}}^{s_{k+1}} f(s,w) \;ds.
\end{equation}
We will approximate the integrals in the right-hand side of (\ref{is}) by using the Taylor expansion of $f(s,w)$ with the center at the right-hand endpoint $s_{k+1}$.\\
We have for all twice differentiable $g(y)$
\begin{equation}\label{L}
g(y+u)-g(y)=u g'(y)+u^2\ds\int\limits_{0}^{1}\theta\ds\int\limits_{0}^{1} g''(\theta\theta_1u+y)\;d\theta_1 d\theta.
\end{equation}
Let $y=s_{k+1}$, $s=s_{k+1}+u$, $u=s-s_{k+1}$. Then, substituting $f(s,w)$ instead of $g(s)$,  (\ref{L}) reads as
\begin{equation}\label{tf}
f(s)=f(s_{k+1})+(s-s_{k+1}) f'(s_{k+1})+(s-s_{k+1})^{2}\ds\int\limits_{0}^{1}\theta\ds\int\limits_{0}^{1}f''\Big(\theta\theta_1(s-s_{k+1})+s_{k+1}\Big)\;d\theta_1 d\theta.
\end{equation}
Let us estimate
\begin{equation*}
R^{(1)}_{k}(s,w):=\ds\int_{0}^{1}\theta\ds\int\limits_{0}^{1} \ds\frac{\partial^2 f}{\partial s^2}\Big(\theta\theta_1(s-s_{k+1})+s_{k+1},w\Big)\;d\theta_1 d\theta,\quad s\in[s_{k},s_{k+1}]. 
\end{equation*}
By (\ref{fs}) 
\begin{equation*}
\ds\frac{\partial^2 f}{\partial s^2}(s,w)=2i\ds\frac{e^{w} e^{2is}-e^{-w}e^{-2is}}{(e^{w} e^{2is}+e^{-w}e^{-2is})^{2}}, 
\end{equation*}
hence we have, by (\ref{p}), the following estimate (and also  estimates of the same type in what follows):
\begin{equation*}
\left|{\ds\frac{\partial^2 f}{\partial s^2}(s,w)}\right|\leq C e^{-\abs{w}}~~{\rm uniformly} ~{\rm in}~s.
\end{equation*}
Therefore
\begin{equation*}
\abs{R^{(1)}_{k}(s,w)}\leq C e^{-\abs{w}}, 
\end{equation*}
$C$ does not depend on $k$ and $w$. 
By (\ref{tf})
\begin{equation}\label{aik}
\begin{array}{lll}
\ds\int\limits_{s_{k}}^{s_{k+1}} f(s,w)\;ds=\ds\int\limits_{s_{k}}^{s_{k+1}} f(s_{k+1},w)\;ds+\ds\int\limits_{s_{k}}^{s_{k+1}} (s-s_{k+1})\ds\frac{\partial f}{\partial s}(s_{k+1},w)\;ds
+\\\\
+\ds\int\limits_{s_{k}}^{s_{k+1}}(s-s_{k+1})^2 R^{(1)}_{k}(s,w)\;ds
=\alpha f(s_{k+1},w)+\ds\frac{\partial f}{\partial s}(s_{k+1},w)\ds\frac{(s-s_{k+1})^2}{2}\Bigg\vert_{s_{k}}^{s_{k+1}}+r^{(1)}_{k}(w),
\end{array}
\end{equation}
where
\begin{equation*}
r^{(1)}_{k}(w)=\ds\int\limits_{s_{k}}^{s_{k+1}}(s-s_{k+1})^2 R^{(1)}_{k}(s,w)\;ds,\quad \abs{r^{(1)}_{k}(w)}\leq \alpha^2 \ds\int\limits_{s_{k}}^{s_{k+1}} C e^{-\abs{w}}\;ds= C e^{-\abs{w}}\alpha^3
\end{equation*}
by (\ref{ncond}).
Hence, by (\ref{is}), (\ref{aik}),
\begin{equation}\label{if}
\ds\int\limits_{-\mu}^{\mu}f(s,w)\;ds=\ds\sum_{k=0}^{2n}\alpha f(s_{k+1},w)-\ds\sum_{k=0}^{2n} \ds\frac{\partial f}{\partial s}(s_{k+1},w) \ds\frac{\alpha^2}{2}+r^{(1)}(w), \quad\abs{r^{(1)}(w)}\leq C e^{-\abs{w}}\alpha^2,
\end{equation} 
where $r^{(1)}(w)=\ds\sum_{k=0}^{2n} r^{(1)}_{k}(w)$. 
Thus, by (\ref{Pn}),
\begin{equation}\label{st1}
\ln P_{n}(w)=\ds\sum_{k=0}^{2n} f(s_{k+1},w)=\ds\frac{1}{\alpha}\ds\int\limits_{-\mu}^{\mu} f(s,w)\;ds+\ds\frac{\alpha}{2} \ds\sum_{k=0}^{2n} \ds\frac{\partial f}{\partial s}(s_{k+1},w)+O(\alpha e^{-\abs{w}}).
\end{equation}
Similarly to the above, we obtain the asymptotics of the sum in the right-hand side of (\ref{st1}) in terms of the integral $\ds\int\limits_{-\mu}^{\mu} \ds\frac{\partial f}{\partial s}(s,w)\;ds$.\\ 
We have
\begin{equation*}
\ds\frac{\partial f}{\partial s}(s,w)=\ds\frac{\partial f}{\partial s}(s_{k+1},w)+(s-s_{k+1}) R^{(2)}_{k}(s,w),
\end{equation*}
where
\begin{equation*}
R_k^{(2)}=\int_0^1{\ds\frac{\partial^2 f}{\partial s^2}\big(\theta s+(1-\theta)s_{k+1},w\big)}\,ds, 
\end{equation*}
and hence
\begin{equation*}
 \abs{R^{(2)}_{k}(s,w)}\leq Ce^{-\abs{w}}.
 \end{equation*}
Integrating over $[s_k,s_{k+1}]$, we obtain
\begin{equation*}
\begin{array}{lll}
\ds\int\limits_{s_{k}}^{s_{k+1}} \ds\frac{\partial f}{\partial s}(s,w)\;ds&=&\ds\int\limits_{s_{k}}^{s_{k+1}} \ds\frac{\partial f}{\partial s}(s_{k+1},w)\;ds+\ds\int\limits_{s_{k}}^{s_{k+1}} (s-s_{k+1})R^{(2)}_{k}(s,w)\;ds\\\\
&=& \ds\frac{\partial f}{\partial s}(s_{k+1},w)\alpha+r^{(2)}_{k}(w),\quad \abs{r^{(2)}_{k}(w)}\leq C\alpha^2  e^{-\abs{w}}. 
\end{array}
\end{equation*}
Hence,
\begin{equation}\label{if'}
\ds\int\limits_{-\mu}^{\mu} \ds\frac{\partial f}{\partial s}(s,w)\;ds=\ds\sum_{k=0}^{2n} \ds\int\limits_{s_{k}}^{s_{k+1}} \ds\frac{\partial f}{\partial s}(s,w)\;ds=\ds\alpha\sum_{k=0}^{2n} \ds\frac{\partial f}{\partial s}(s_{k+1},w) +r^{(2)}(w),
\end{equation}
where
\begin{equation*}
\abs{r^{(2)}(w)}\leq \ds\sum_{k=0}^{2n} C\alpha^2 e^{-\abs{w}}=2n C \alpha^2 e^{-\abs{w}}\leq  C_1\alpha e^{-\abs{w}} ~~{\rm by}~(\ref{lam_{n}^2}). 
\end{equation*}
By (\ref{if'}),
\begin{equation}\label{sf'}
\alpha\ds\sum_{k=0}^{2n} \ds\frac{\partial f}{\partial s}(s_{k+1},w)=\ds\int\limits_{-\mu}^{\mu} \ds\frac{\partial f}{\partial s}(s,w)ds+r^{(2)}(w),\quad \abs{r^{(2)}(w)}\leq C_1\alpha e^{-\abs{w}}.
\end{equation}
Substituting the value of the sum in (\ref{sf'}) into (\ref{st1}), we obtain
\begin{equation}\label{st2}
\ln P_{n}=\ds\frac{1}{\alpha} \ds\int\limits_{-\mu}^{\mu}f(s,w)\;ds+\ds\frac{1}{2}\ds\int\limits_{-\mu}^{\mu} \ds\frac{\partial f}{\partial s}(s,w)\;ds+O(\alpha e^{-\abs{w}}).
\end{equation}
The last formula is the representation of the kernel $P_n$ in terms of simple integrals which we were looking for. It can be simplified further as follows:
\begin{equation}\label{if''}
\ds\int\limits_{-\mu}^{\mu} \ds\frac{\partial f}{\partial s}(s,w)\;ds=f(\mu,w)-f(-\mu,w)=\ln\ds\frac{\cot\Big(\ds\frac{\pi}{4}-\ds\frac{iw}{2}+\mu\Big)}{\cot\Big(\ds\frac{\pi}{4}-\ds\frac{iw}{2}-\mu\Big)},      
\end{equation}
and it is easy to see that 
\begin{equation}\label{cot/cot}
\ds\frac{\cot\Big(\ds\frac{\pi}{4}-\ds\frac{iw}{2}+\mu\Big)}{\cot\Big(\ds\frac{\pi}{4}-\ds\frac{iw}{2}-\mu\Big)}=\ds\frac{\cosh w-\sin2\mu}{\cosh w+\sin2\mu}.
\end{equation}
Hence
\begin{equation}\label{st3}
\ln P_{n}=\ds\frac{1}{\alpha} \ds\int\limits_{-\mu}^{\mu} f(s,w)\;ds+\ds\frac{1}{2}\ln \ds\frac{\cosh w-\sin2\mu}{\cosh w+\sin2\mu}+O(\alpha e^{-\abs{w}}).
\end{equation}
Let us rewrite the last formula in a form close to the WKB-solutions (\ref{psi_n}).\\ 
Denote
\begin{equation}\label{1.5'}
S(w):=\ds\frac{1}{i}\ds\int\limits_{-\mu}^{\mu} f(s) ds,\qquad A(w):=\Big(\ds\frac{\cosh w-\sin2\mu}{\cosh w+\sin2\mu}\Big)^{1/2},
\end{equation}
where the square root is arithmetic.
The function $S(w)$ will play the role of the phase function in the WKB solutions (\ref{psi_n}) and hence should have the form of a primitive of the analog of the coordinate function $q$ (see (\ref{Fi}) below). In order to achieve this goal, we calculate its derivative $S'(w)$.
We have by (\ref{fs}), (\ref{cot/cot}), (\ref{if''}) that
\begin{equation}\label{S'}
\begin{array}{lll}
S'(w)&=&-\ds\frac{1}{2}\ds\int\limits_{-\mu}^{\mu} \ds\frac{\partial f}{\partial s}(s,w) ds=-\ds\frac{1}{2}\Big(f(\mu,w)-f(-\mu,w)\Big)
=-\ds\frac{1}{2}\ln\ds\frac{\cot\Big(\ds\frac{\pi}{4}-\ds\frac{iw}{2}+\mu\Big)}{\cot\Big(\ds\frac{\pi}{4}-\ds\frac{iw}{2}-\mu\Big)}\\\\
&=&-\ds\frac{1}{2}\ln\Big(\ds\frac{\cosh w-\sin2\mu}{\cosh w+\sin2\mu}\Big)=-\ln A(w).
\end{array}
\end{equation}
Let us calculate $S(0)$. We have by (\ref{1.5'})
\begin{align*}
& S(0)=\ds\frac{1}{i}\ds\int\limits_{-\mu}^{\mu} \ln\cot\Big(\ds\frac{\pi}{4}+s\Big)ds. 
\end{align*}
Since
\begin{align*}
& \ds\int\limits_{-\mu}^{0}\ln\cot\Big(\ds\frac{\pi}{4}+s\Big)ds=
\ds\int\limits_{0}^{\mu}\ln\cot\Big(\ds\frac{\pi}{4}-s\Big)ds=-\ds\int\limits_{0}^{\mu}\ln \cot\Big(\ds\frac{\pi}{4}+s\Big)\,ds,
\end{align*}
$\Bigg($because $\cot\Big(\ds\frac{\pi}{4}-s\Big)=\ds\frac{1}{\cot\Big(\ds\frac{\pi}{4}+s\Big)}
$$\Bigg)$, we have
\begin{equation*}
S(0)=\ds\frac{1}{i}\ds\int\limits_{0}^{\mu}\Bigg(\ln \cot\Big(\ds\frac{\pi}{4}+s\Big)-\ln \cot\Big(\ds\frac{\pi}{4}+s\Big)\Bigg)ds=0,
\end{equation*}
and by (\ref{S'}) we come to the representation of $S$ as a primitive:
\begin{equation}\label{S(w)}
S(w)=-\ds\frac{1}{2}\ds\int\limits_{0}^{w}\ln \ds\frac{\cosh z-\sin2\mu}{\cosh z+\sin2\mu} dz.
\end{equation}
Finally, by (\ref{st3}), (\ref{1.5'}) we obtain the asymptotics of the logarithm of $P_n$:
\begin{equation}\label{lP1}
\ln P_n=\ds\frac{i}{\alpha}S(w)+\ln A(w)+O(\alpha e^{-\abs{w}}),
\end{equation}
where $A(w)$ is given in (\ref{1.5'}).
Note that the functions $S$ and $A$ depend regularly on $\lambda$, $0\leq\lambda<1$; we will sometimes omit this dependence.
Finally, by (\ref{fin1}), (\ref{lP1}), (\ref{1.5'})
\begin{equation*}
\begin{array}{lll}
\varphi_{n}^{(1)}(x,y)&=&\ds\frac{1}{4\pi}\ds\int\limits_{\underrightarrow{\Gamma_{-\alpha}}} e^{-i\rho\sinh (w+i\theta)}\;e^{\ln P_{n}^{(1)}(w)}\;dw
= \ds\frac{1}{4\pi}\ds\int\limits_{\Gamma_{-\alpha}} e^{-i\rho\sinh (w+i\theta)}\;e^{{i}S(w)/{\alpha}}\;A(w)\;dw+\\\\
&+& O(\alpha)=
 \ds\frac{1}{4\pi}\ds\int\limits_{\Gamma_{-\alpha}} e^{-i\rho \sinh(w+i\theta)+{i}S(w)/{\alpha}}\Big(\ds\frac{\cosh w-\sin 2\mu}{\cosh w+\sinh \mu}\Big)^{1/2}\;dw+O(\alpha).
\end{array}
\end{equation*}
Here and everywhere below the $O$-symbol is uniform in $(x,y)\in\Omega$ and $n$ satisfying (\ref{ncond}).

Since $S(w)$ on the real axis is purely real and its derivative is bounded for $w\in \Gamma_{-\alpha}$, the function $A$ is bounded for $w\in\Gamma_{-\alpha}$, the exponential $e^{-i\rho\sinh(w+i\theta)}$ is also bounded for $w\in\Gamma_{-\alpha}$, $0<\theta<\alpha$; hence  
\begin{equation*}
e^{\ln P_{n}^{(1)}(w)}=e^{{i}S(w)/{\alpha}} A(w)\Big(1+O(\alpha e^{-\abs{w}})\Big)
\end{equation*} 
and thus we have proven the following 
\begin{lem}\label{1}
\begin{equation}\label{varphi1}
\varphi_{n}^{(1)}(x,y)=\ds\frac{1}{4\pi} \ds\int\limits_{\underrightarrow{\Gamma_{-\alpha}}} e^{-i\rho \sinh(w+i\theta)}\;e^{{iS(w)}/{\alpha}} A(w) dw+O(\alpha).
\end{equation} 
\end{lem} 
 \begin{obs}\label{rem1}
 The integral (\ref{varphi1}) converges since $0\leq y\leq x\tan\alpha$.
In fact, on $\Gamma_{-\alpha}$, for $w_1\in \R$,
\begin{equation*}
\cosh(w_1-i\alpha)=\cosh w_1\cos\alpha-i\sinh w_1\sin\alpha
\end{equation*}
and hence
\begin{equation*}
\begin{array}{lll}
\abs{e^{-ix \sinh (w_1-i\alpha)}\;e^{y\cosh(w_1-i\alpha)}}=e^{-x\cosh w_1\sin\alpha+y\cosh w_1\cos\alpha}\longrightarrow 0\quad \text{for}~ x>0,~{\rm and}~w_1\longrightarrow \pm\infty
\end{array}
\end{equation*}
because we have
\begin{equation*}
 y\cosh w_1\cos\alpha<x\tan\alpha\cosh w_1\cos\alpha=x\sin\alpha\cosh w_1 \quad \text{for} ~0<y<x\tan\alpha.
\end{equation*}
\end{obs}
\begin{obs}\label{rem2}
The error term $O(\alpha)$ in (\ref{varphi1}) can in its turn be calculated explicitly by calculating the corrections in (\ref{st3}); we restrict ourselves only to the leading term.
\end{obs}

\subsection{Asymptotics of $\varphi_{n}^{(2)}$}
We will show that the integral over $\Gamma_{-\pi}$ in (\ref{phn2}) can be reduced to an integral over $\Gamma_{-\alpha}$.
Changing the variable: $w'=-w-i\pi,~w=-w'-i\pi,~w\in\Gamma_{-\pi}, w'\in\R$ in (\ref{phn2}), we obtain
from  (\ref{Ph1}) 
\begin{equation*}
\varphi_{n}^{(2)}(x,y)=\ds\frac{1}{4\pi}\ds\int\limits_{\underrightarrow{\Gamma_{0}}} \; e^{-i\rho\sinh( w-i\theta)}P_n(-w'-i\pi) (-dw')
=\ds\frac{1}{4\pi}\ds\int\limits_{\underrightarrow{\Gamma_{0}}} \; e^{-i\rho\sinh (w-i\theta)}Q_n(w) dw,
\end{equation*}
where 
\begin{equation}\label{Pn2}
Q_{n}(w)=\prod_{k=0}^{2n} \cot\Big(\ds\frac{\pi}{4}-\ds\frac{iw}{2}+\alpha\big(n-k-\ds\frac{1}{2}\big)\Big), \qquad \Gamma_0=\{\Im w=0\}.
\end{equation}
The contour $\underrightarrow{\Gamma_{0}}$ can be transformed into $\underrightarrow{\Gamma_{-\alpha}}$ by the Cauchy theorem and we obtain  (cf. (\ref{fin1}))
\begin{equation}\label{phn21}
\varphi_{n}^{(2)}(x,y)=\ds\frac{1}{4\pi}\ds\int\limits_{\underrightarrow{\Gamma_{-\alpha}}}\; e^{-i\rho\sinh (w-i\theta)}Q_n(w) dw.
\end{equation}
Denote $\beta:=\ds\frac{\pi}{4}-\ds\frac{iw}{2}$. We have
\begin{equation*}
\begin{array}{lll}
\cot\Big(\beta-\ds\frac{iw}{2}\Big)=\ds\frac{\sin2\beta-i\sinh w}{\cosh w-\cos2\beta}.
\end{array}
\end{equation*}
Similarly to the above, 
by definition (\ref{Pn2}) of $Q_{n}(w)$ we have (after the change $k\to 2n-k$) 
\begin{equation}\label{P_n2}
\ln Q_{n}(w)=\ds\sum_{k=0}^{2n} \ln \cot\Big(\ds\frac{\pi}{4}-\ds\frac{iw}{2}+\alpha\big(-n+k-\ds\frac{1}{2}\big)\Big)=\ds\sum_{k=0}^{2n} f(s_{k},w).
\end{equation}
Recall that $\mu=\alpha\left(n+{1}/{2}\right)$, $f(s)$ is given by (\ref{fs}) and $s_{k}$ is given by (\ref{Pn}).
Then, using the Taylor expansion with center at the left-hand end points $s_{k}$, we have
\begin{equation}\label{-mintm}
\ds\int\limits_{-\mu}^{\mu} f(s,w)\;ds=\ds\sum_{k=0}^{2n} \ds\int\limits_{s_{k}}^{s_{k+1}} f(s,w)\;ds=\ds\sum_{k=0}^{2n}\Bigg(f(s_{k},w)\cdot\alpha+\ds\frac{\partial f}{\partial s}(s_{k},w)\ds\frac{\alpha^2}{2}+O(\alpha^3 e^{-\abs{w}})\Bigg)
\end{equation}
since
\begin{equation*}
f(s,w)=f(s_{k},w)+\ds\frac{\partial f}{\partial s}(s_{k},w)(s_{k+1}-s_{k})+O(\alpha^2 e^{-\abs{w}}),\quad  s\in[s_{k},s_{k+1}].
\end{equation*}
Hence by (\ref{P_n2}), (\ref{-mintm}), (\ref{S'}), similarly to the previous subsection, we obtain 
\begin{equation}\label{lpn2}
\ln Q_{n}(w)=\ds\frac{1}{\alpha} \ds\int\limits_{-\mu}^{\mu} f(s,w)\;ds-\ds\frac{1}{2}\ds\int\limits_{-\mu}^{\mu} \ds\frac{\partial f}{\partial s}(s,w)\;ds+O(\alpha e^{-\abs{w}})=\ds\frac{iS(w)}{\alpha}-\ln A+O(\alpha e^{-\abs{w}}),
\end{equation}
and
\begin{equation*}
Q_{n}(w)=e^{i\frac{S(w)}{\alpha}}\;A^{-1}(w)+O(\alpha e^{-|w|},
\end{equation*}
where $A$ is defined in (\ref{1.5'}).  Therefore, by (\ref{phn21}), (\ref{lpn2}),
\begin{equation*}
\varphi_{n}^{(2)}(x,y)=\ds\frac{1}{4\pi}\ds\int\limits_{\underrightarrow{\Gamma_{-\alpha}}} e^{-i\rho\sinh(w-i\theta)+{iS(w)}/{\alpha}}\;A^{-1}(w)\;dw +O(\alpha).
\end{equation*}
 Thus we come to the following 
\begin{lem}\label{2}
\begin{equation}\label{phin2}
\begin{array}{lll}
\varphi_{n}^{(2)}(x,y)
=\ds\frac{1}{4\pi}\ds\int\limits_{\underrightarrow{\Gamma_{-\alpha}}}\;e^{-ix\sinh w-y\cosh w}\; e^{{iS(w)}/{\alpha}} \;A^{-1}(w)\;dw+O(\alpha).
\end{array}
\end{equation}
\end{lem}
Using Lemmas\;\ref{1} and  \ref{2}, we finally obtain 
\begin{equation}\label{fin}
\begin{array}{lll}
\varphi_{n}(x,y)=\Big(\varphi_{n}^{(1)}+\varphi_{n}^{(2)}\Big)(x,y)
=\ds\frac{1}{4\pi}\ds\int\limits_{\underrightarrow{\Gamma_{-\alpha}}}\;e^{-ix\sinh w}\;e^{{i}S(w)/{\alpha}}\Big(e^{y\cosh w} A+e^{-y\cosh w} A^{-1}\Big) dw+O(\alpha).
\end{array}
\end{equation}

\subsection{WKB-representation of the solutions}
Rewrite the expression in the parenthesis in (\ref{fin}) as follows
\begin{equation}\label{B(w)}
e^{y\cosh w} A+e^{-y\cosh w} A^{-1}=\left(A+A^{-1}\right)\cosh(y\cosh w)
+\left(A-A^{-1}\right)\sinh(y\cosh w).
\end{equation}
Further,
\begin{equation*}
 A+A^{-1}= \ds\frac{2\cosh w}{\left(\cosh^2w-\lambda^2\right)^{1/2}},\qquad
A-A^{-1}= \ds\frac{2\lambda}{\left(\cosh^2w-\lambda^2\right)^{1/2}}.
\end{equation*}
Substituting (\ref{B(w)}) in (\ref{fin}), we obtain  
\begin{equation}\label{phiw}
\varphi_{n}(x,y)=\ds\frac{1}{4\pi}\ds\int\limits_{\underrightarrow{\Gamma_{-\alpha}}} e^{\frac{i}{\alpha}(-\xi\sinh w+S(w,\lambda))}\; B(w,y,\lambda) dw+O(\alpha),
\end{equation}
where $\xi=\alpha x$ is the ``slow'' variable and
\begin{equation*}
B(w,y)=\ds\frac{2\cosh w}{\big(\cosh^2 w-\lambda^2\big)^{1/2}} \cosh(y\cosh w)+\ds\frac{2\lambda}{\big(\cosh^2 w-\lambda^2\big)^{1/2}}\sinh(y\cosh w).
\end{equation*}

In order to reduce formula (\ref{phiw}) to the WKB solutions (\ref{psi_n}), we perform the change of the variable
 $$\sinh w=p,\quad\text{ i.e.,} \quad w(p)=\operatorname{arsinh}p,$$
  in (\ref{fin}). Thus we come to the following
\begin{teo}
The Ursell modes (\ref{n>0}) for $n$ satisfying (\ref{ncond}) have the form
\begin{align}\label{phi(x)}
& \varphi_{n}(x,y)=\ds\frac{1}{2\pi}\ds\int\limits_{C_{\alpha}} e^{-{i\xi p}/{\alpha}+{i}\Phi(p)/{\alpha}}\; b(p,y,\lambda)\; dp+O(\alpha),
\end{align}
where
\begin{equation}\label{B(p,y)}
 b(p,y,\lambda)=\ds\frac{1}{2\tau}B\Big(w(p), y,\lambda\Big)=\ds\frac{\cosh y\tau-\lambda\tau^{-1}\sinh y\tau}{\tau\Big(1-\lambda^2/\tau^2\Big)^{1/2}},\quad \tau=\sqrt{1+p^2},\quad \Phi(p,\lambda)= S\Big(w(p,\lambda)\Big),
\end{equation}
and the contour $C_{\alpha}$ is the image of $\Gamma_{-\alpha}$ under the mapping $p=\sinh w$, that is, $C_{\alpha}$ is the hyperbola 
$$
\ds\frac{(\Im p)^2}{\sin^2\alpha}-\ds\frac{(\Re p)^2}{\cos^2\alpha}=1, \quad\Im p<0. 
$$
\end{teo}
\begin{obs}\label{rWKB}
This theorem states that, as mentioned in the Introduction, the Ursell modes have the form of the WKB-solutions (\ref{psi_n}) in the momentum representation. The contour $C_{\alpha}$, as we will see below (cf. also \cite{Zh95}), can be deformed into a closed contour on the plane $p\in\C$. 
\end{obs}

\section{Asymptotics of the WKB-representation}
\begin{figure}[htbp]
\centering
\includegraphics[scale=0.27]{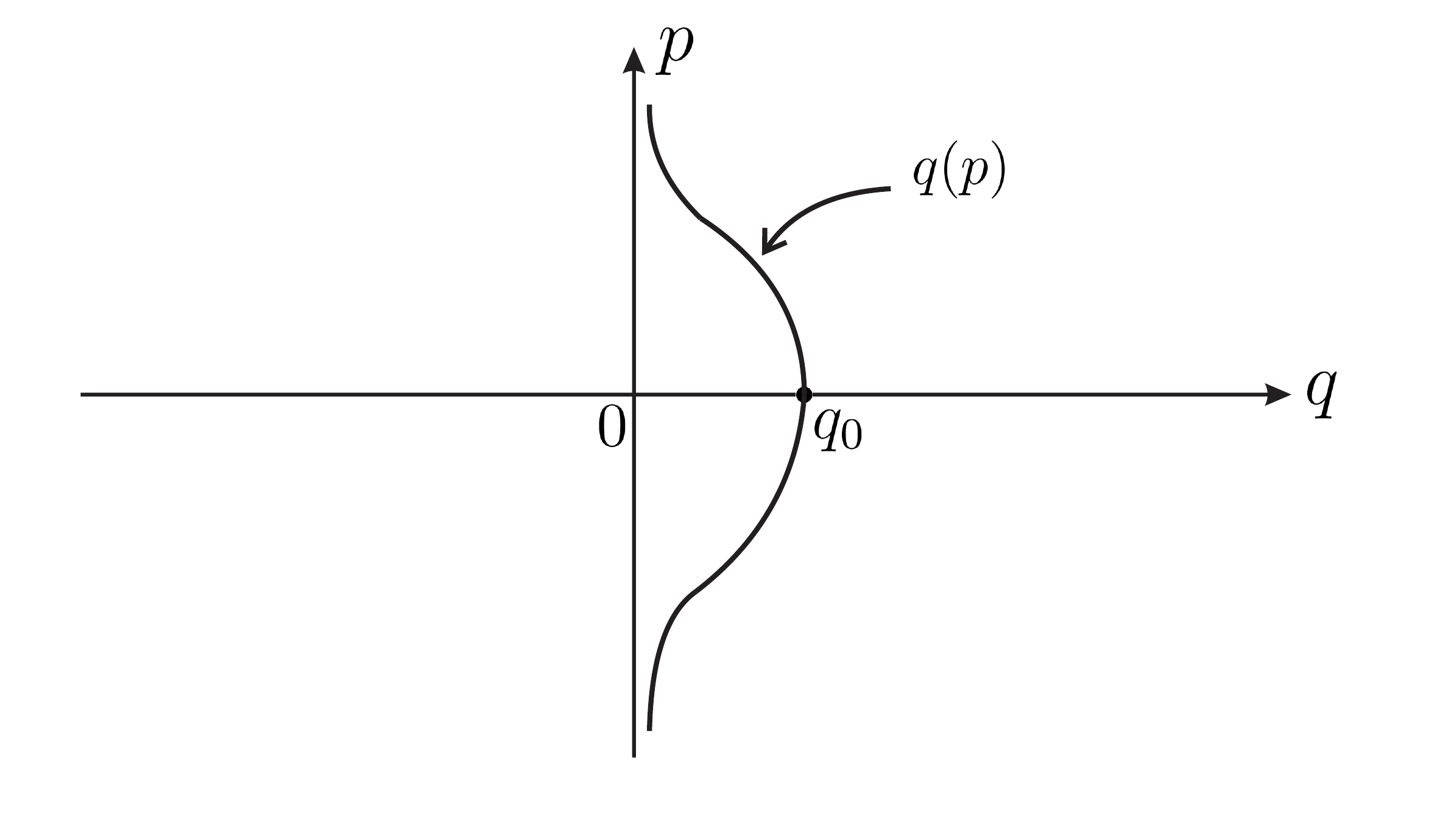}\caption{The graph of $q(p)$}\label{qq}
\end{figure}

Clearly, $S(w)$ is real for real $w$ by (\ref{S(w)}) and hence, for $w\in\Gamma_{-\alpha}$, the imaginary part of $S$ is of order of $O(\alpha)$.
The integral (\ref{phiw}) has the standard form of a rapidly oscillating integral normally treated by means of the stationary phase method and its generalizations.
For real $w$, the graph of the function $S'(w)$ is similar to the one shown in Fig.\;\ref{qq}, where the coordinate $p=\sinh w$ corresponds to the variable $w$ and $q(p) =S'(w(p))/\cosh w(p)$.

 In what follows the coordinate $p$ (the classical momentum) will play a crucial role (see (\ref{phi(x)})). For $\xi\geq q_{0}+\delta_1$, where $q_0=\operatorname{artanh}\lambda$, with any $\delta_1>0$, the derivative of the phase $-\xi\sinh w+S(w)$ is bounded away from $0$. Indeed, the stationary points of the phase are solutions of the equation $\xi=S'(w)/\cosh w$, and, if $\xi\geq q_0+\delta_1$, this equation does not have solutions for real $w$ (and hence for $w\in \Gamma_{-\alpha}$). Hence it is possible to integrate by parts in (\ref{phiw}) any number of times and the integral in (\ref{phiw}) is $O(\alpha^{\infty})$. Thus in what follows we will be interested in the asymptotics of (\ref{phiw}) only for $\xi\leq q_0+\delta_1$, $\delta_1>0$.\\   
\setcounter{equation}{0}
\subsection{
The closure of the integration contour}

\begin{figure}[htbp]
\centering
\includegraphics[scale=0.27]{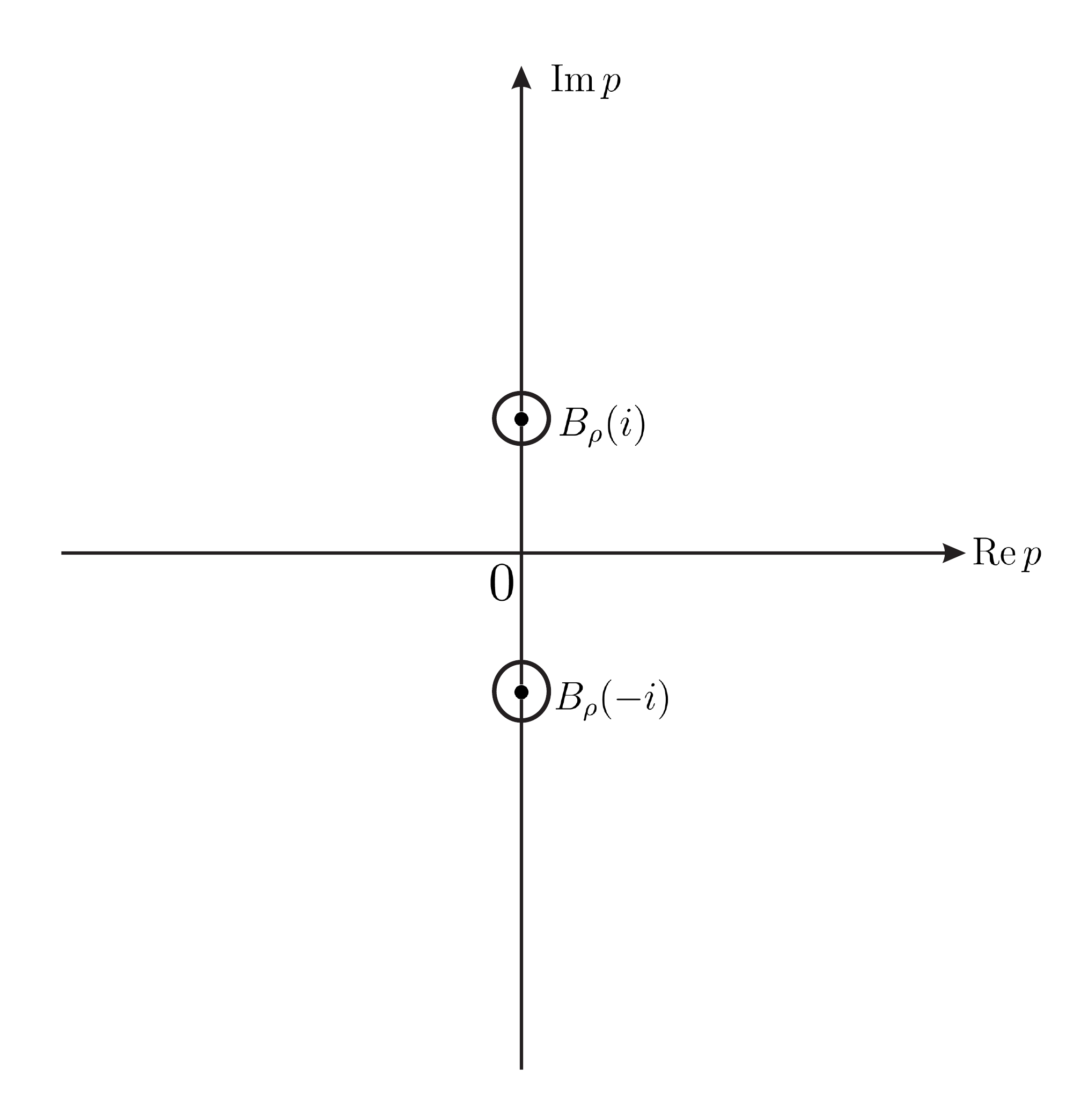}
\caption{The domain of $q(p)$}\label{Bp}
\end{figure}

Let us analyze the analytic properties of the functions entering the WKB approximations (\ref{phi(x)}) (cf. \cite{Zh92,SunShen}).
Let us calculate the values of  $ \Phi(p,\lambda)=S\big(w(p)\big)$. We have by (\ref{S(w)})
\begin{equation*}
\Phi(p,\lambda)=\ds\frac{1}{2}\ds\int\limits_{0}^{w(p)}\ln\ds\frac{\cosh w'+\sin2\mu}{\cosh w'-\sin2\mu} dw',\quad \Im p<0.
\end{equation*}
Let us make the change of the variable $\sinh w'=p'$ in the last integral. We have $dw'=\ds\frac{dp'}{\sqrt{1+p'^{2}}}$, $\cosh w'=\sqrt{1+p'^{2}}$.
Hence 
\begin{align}\label{Fi}
& \Phi(p)=\ds\frac{1}{2}\ds\int\limits_{0}^{p}\ln\ds\frac{\sqrt{1+p'^{2}}+\sin2\mu}{\sqrt{1+p'^{2}}-\sin2\mu}\;\ds\frac{dp'}{\sqrt{1+p'^{2}}}=\ds\int\limits_{0}^{p} q(p) dp.
\end{align}
where (see (\ref{S'}))
\begin{equation}\label{q}
q(p)=\frac{S'(w(p))}{\cosh w(p)}=\ds\frac{1}{2\tau}\ln\Bigg(\ds\frac{\tau+\sin2\mu}{\tau-\sin2\mu}\Bigg)=\ds\frac{1}{\tau}\operatorname{artanh}\ds\frac{\lambda}{\tau},\quad \tau=\sqrt{1+p^2}.
\end{equation}
Obviously, $q(p)$ given in (\ref{q}) is analytic on the plane $p\in\C$ outside small neighborhoods of the points $p=\pm i$ such that $\lambda/\tau<1$; see Fig.\;\ref{Bp}.

\begin{figure}[htbp]
\centering
\includegraphics[scale=0.27]{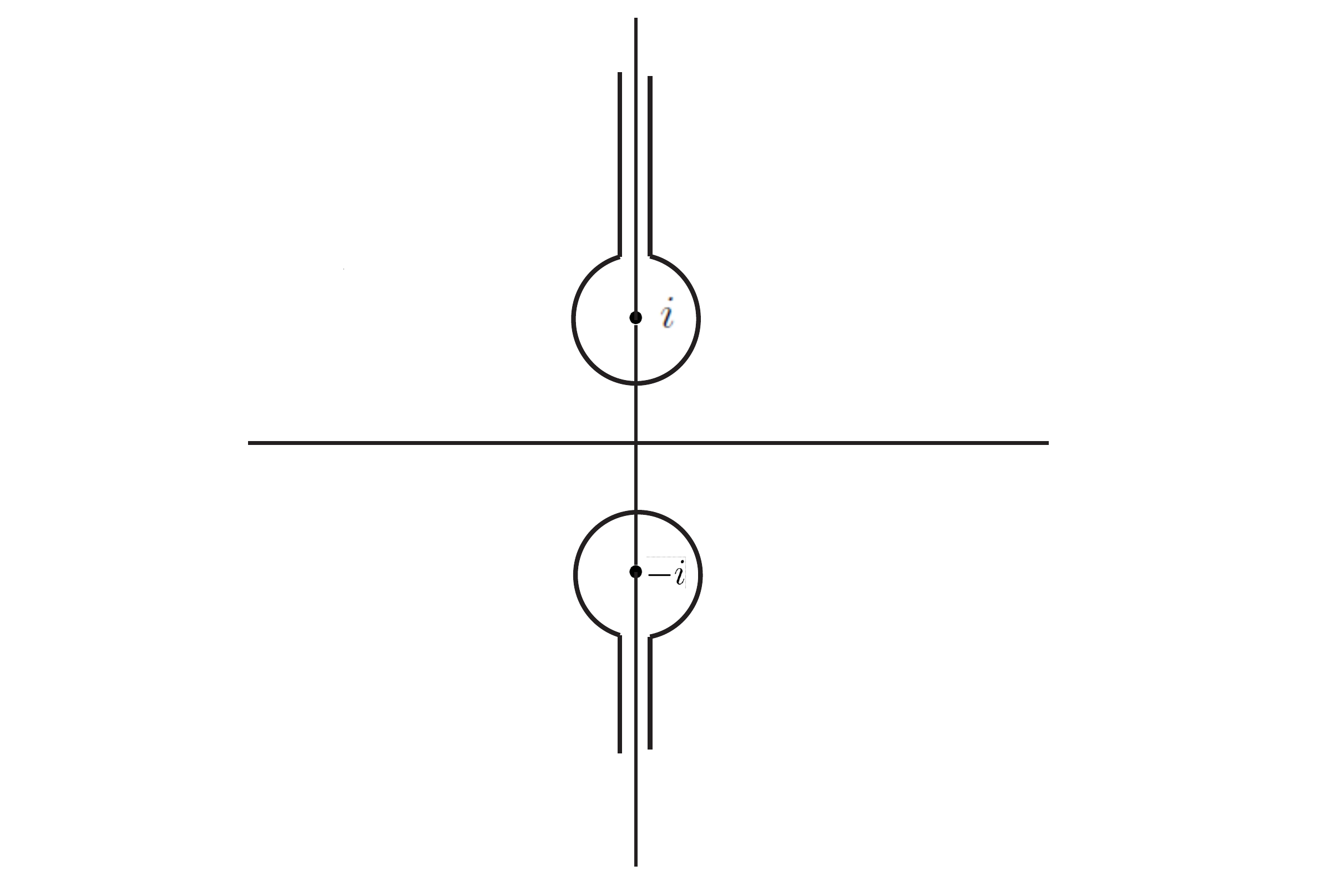}
\caption{The domain of $\Phi(p)$}\label{ii3}
\end{figure}

 Hence, by (\ref{Fi}), the function $\Phi(p)$ is analytic on the plane $\C$ outside small neighborhoods of the points $\pm i$ and with cuts along the rays $(-i\infty,-i)\cup (i,i\infty)$; see Fig. \ref{ii3}.

 Let us calculate the values of $\Phi(p)$ for $p=\pm\infty$.
We have by(\ref{Fi})
\begin{align*}
& \Phi(\pm\infty, \lambda)=\ds\int\limits_{0}^{\pm \infty}\ds\frac{1}{\sqrt{1+p^2}}\text{artanh}\ds\frac{\lambda}{\sqrt{1+p^2}} \;dp,\quad \lambda<1.
\end{align*}
Differentiating with respect to $\lambda$, we obtain
\begin{align*}
& \ds\frac{\partial\Phi}{\partial \lambda}(\pm\infty, \lambda)=\ds\int\limits_{0}^{\pm \infty}\ds\frac{1}{p^2+1-\lambda^2} \;dp=\pm\ds\frac{\pi}{2\sqrt{1-\lambda^2}}
\end{align*}
Hence by (\ref{lam_{n}^2})
\begin{align*}
& \Phi(\pm\infty, \lambda)=\pm\ds\frac{\pi}{2}\arcsin\lambda=\pm\ds\frac{\pi}{2}\Big(2n+1\Big)\alpha
\end{align*}   
and
\begin{align}\label{quc}
& \Phi(\infty)-\Phi(-\infty)=\int_{-\infty}^{\infty}q(p)\,dp=\pi(2n+1)\alpha.
\end{align}
Since $q(p)\sim\lambda/\tau^2$ for large $\tau$, we have 
\begin{equation}\label{coro}
\int_{-\infty}^{\infty}q(p)\,dp=\oint_{{C^{-}}}q(p)\,dp=\pi(2n+1)\alpha,
\end{equation}
where the contour $C^{-}$ is any closed simple contour encircling $B_{\rho}(-i)$; the index $``-''$ means the clockwise direction. 

\begin{figure}[htbp]
\centering
\includegraphics[scale=0.27]{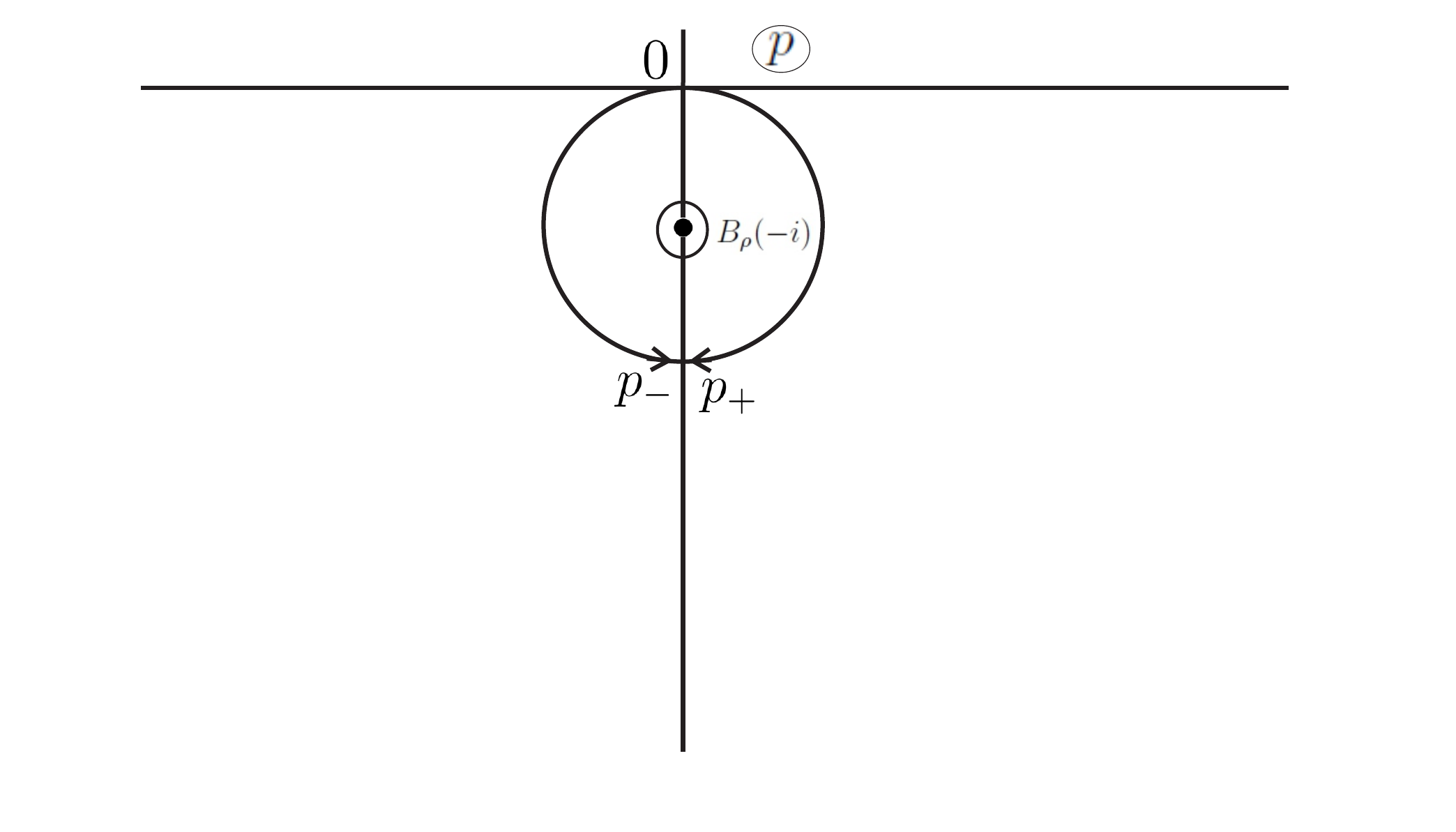}
\caption{ Calculation of the jump of the function $\Phi(p)$}\label{pp}.
\end{figure}

Let us calculate the jump of $\Phi(p)$ across the cut $(-i\infty,-i-i\rho)$. We have (see Fig. \ref{pp})
\begin{equation}\label{jump phi}
\Phi(p_{+})-\Phi(p_{-})=\oint_{{C^{-}}}q(p)\,dp=\pi(2n+1)\alpha.
\end{equation}

\begin{figure}[htbp]
\centering
\includegraphics[scale=0.27]{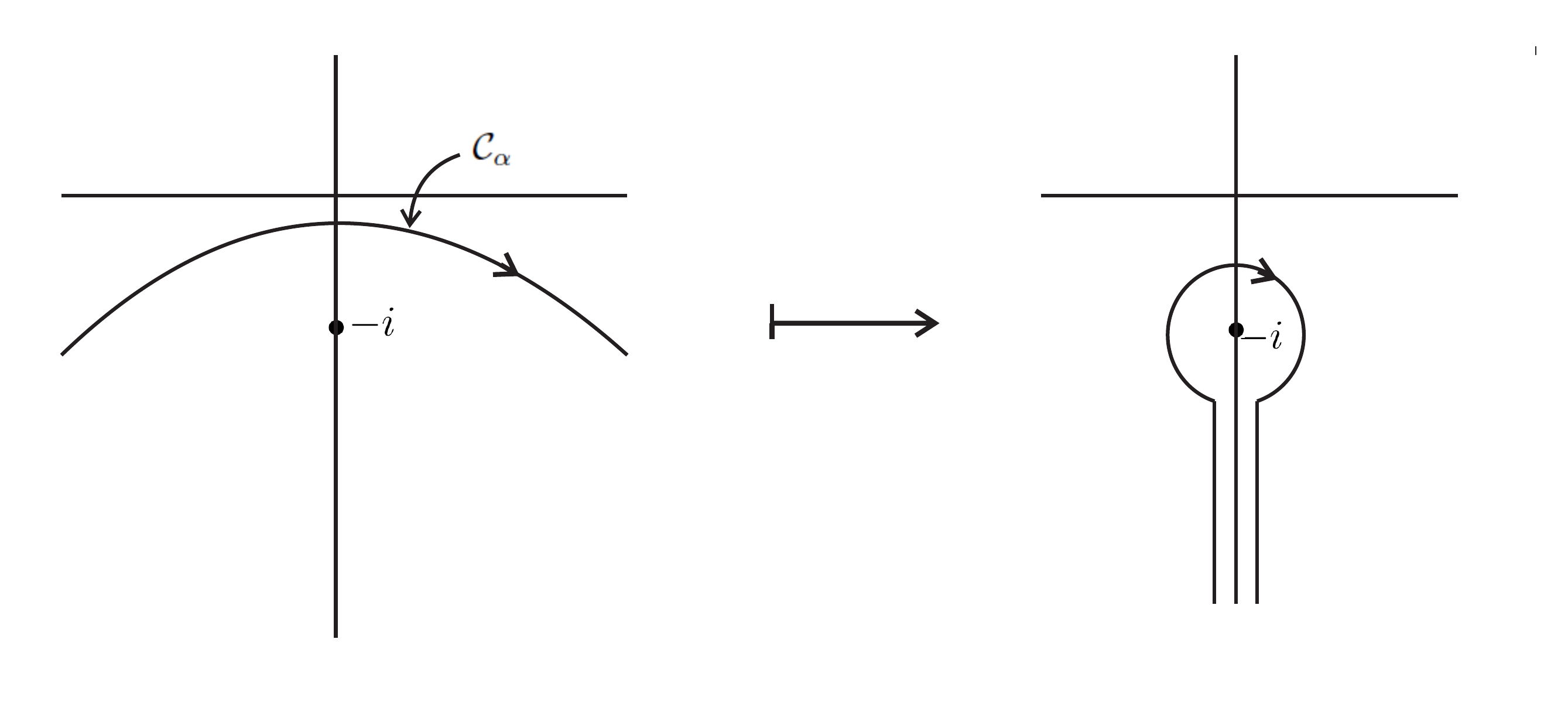}
\caption{The closure of the contour $\mathcal{C}_{\alpha}$}\label{Oj2}
\end{figure}

\begin{figure}[htbp]
\centering
\includegraphics[scale=0.27]{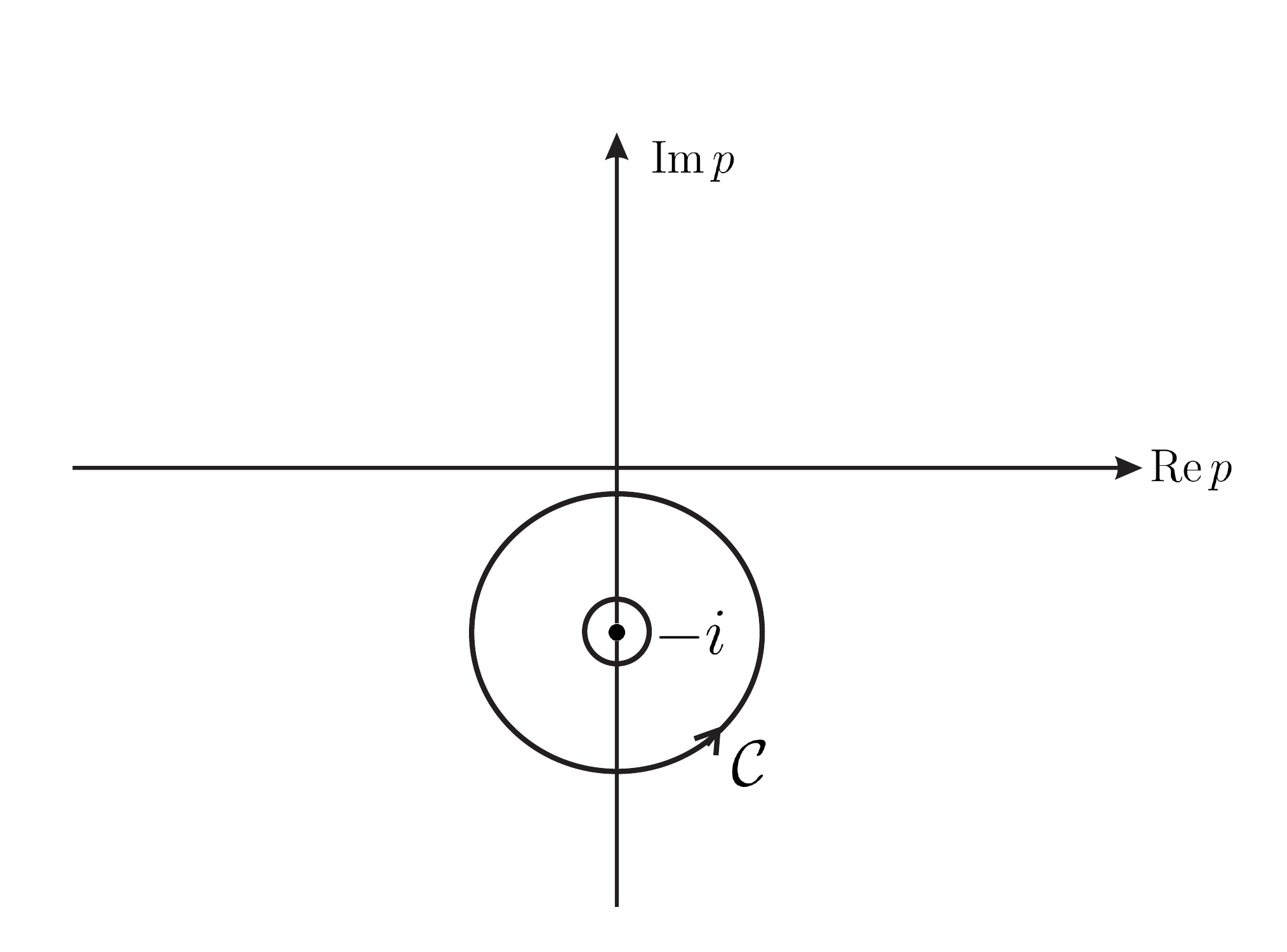}
\caption{The contour $\mathcal{C}$}\label{C1}
\end{figure}

Now we change the contour of integration $\mathcal{C}_\alpha$ in (\ref{phi(x)}) to the contour  in the right-hand side of Fig. \ref{Oj2}. This is possible since the growth of the function $b$ in (\ref{phi(x)}) as $\tau\to\infty$ is suppressed by the exponential $\exp(-i\xi p/\alpha)$ (cf. Remark \ref{rem1}) and $\Phi(p)$ tends to a constant as $p\to\infty$ ( see (\ref{S(w)})  and (\ref{B(p,y)})). We show that this contour reduces to the closed simple contour $\mathcal{C}$ in Fig. \ref{C1}.

We  prove that
\begin{equation}\label{I2}
\ds\int\limits_{-i\infty}^{-iR}b(p,y) e^{-{ip\xi}/{\alpha}+{i}{\Phi(p)}/\alpha}\;dp+\ds\int\limits_{-iR}^{-i\infty}b(p,y) e^{-{ip\xi}/{\alpha}+{i}\Phi(p)/\alpha}\;dp=0.
\end{equation}
Indeed, the  function $b(p,y)$ is analytic in $p$ in the domain shown in Fig.\;\ref{Bp} since both functions in the numerator of (\ref{phi(x)}) have Taylor expansions in even powers of $\tau$. Hence, due to the fact that the function $\tau$ acquires the factor  $\exp(i\pi)$ when crossing the ray $(-i\infty,-i-i\rho)$, we have  
\begin{equation}\label{jump b}
b(p_{-},y)=b(p_{+},y) e^{i\pi}.
\end{equation}
Taking into account formula (\ref{jump phi}), we obtain
\begin{equation*}
b(p_{-},y) e^{{-ip_{-}\xi}/{\alpha}+{i}\Phi(p_{-})/\alpha}=b(p_{+},y) e^{{-ip_{+}\xi}/{\alpha}+{i}\Phi(p_{+})/\alpha}.
\end{equation*}
Therefore, the integrand in (\ref{phi(x)}) is analytic on the plane $\C$ outside small neighborhoods of $\pm i$. 
Hence (\ref{I2}) holds and 
\begin{equation}\label{2.8}
\varphi_{n}(x,y)=\ds\frac{1}{2\pi}\ds\oint\limits_{\mathcal{C}^{-}} e^{{-ip\xi}/{\alpha}+{i\Phi(p)}/{\alpha}}\;b(p,y) dp+	O(\alpha),
\end{equation}
where $\mathcal{C}$ is a circle with center at $p=-i$, the index ``$-$'' means the clockwise orientation of the contour (see Fig. \ref{C1}).

\subsection{The asymptotics}

By (\ref{2.8}), the derivation of the asymptotics of (\ref{n>0}) reduces to the calculation of the integrals
\begin{equation}\label{UB(p)}
-\ds\frac{1}{2\pi}\ds\oint\limits_{\mathcal{C}} b(p,y,\lambda) e^{-{ip\xi}/{\alpha}+{i}\Phi(p)/\alpha}\;dp.
\end{equation}

Clearly, the asymptotics of (\ref{UB(p)}) depends on the behaviour of the stationary points of the phase function $-p\xi+\Phi(p)$ in (\ref{UB(p)}). Note that these stationary points depend on the parameters $\xi$ and $\lambda$ and are solutions with respect to $p$ of the equation (see (\ref{Fi}) and Fig. \ref{qq})
\begin{equation}\label{3-5-1}
-\xi+q(p)=0.
\end{equation}

Note that, by (\ref{q}) the function $q(p)$ has the form 
\begin{equation*}
q(p)=zh(\lambda z), ~~{\rm where}~z=\lambda/(1+p^2),
\end{equation*}
$h$ is analytic at the origin and $h(0)=1$. Consider the equation (\ref{3-5-1}) in terms of $z$:
\begin{equation}\label{lambda z}
-\xi+zh(\lambda z)=0.
\end{equation}
Denote the solution of (\ref{lambda z}) with respect to 
$z$ by $z_0(\xi,\lambda)$; clearly, by the Implicit Function Theorem, and the definition of $q(p)$ (see (\ref{q})), $z_0(\xi,\lambda)$ is analytic in $\xi,\lambda$ and $z_0(\xi,\lambda)=\xi+O(\lambda \xi)$. Hence, the solutions of (\ref{3-5-1}) are given by
\begin{equation}\label{3-5-3}
p=\pm p_0(\xi,\lambda)=\pm \sqrt{\ds\frac{\lambda}{z_0(\xi,\lambda)}-1},
\end{equation}
where the branch of the square root is arithmetic for sufficiently small real $\xi$ (see Fig.\;\ref{qq}).

Now we are ready to obtain the asymptotics of the expression (\ref{UB(p)}) which differs from $\varphi_{n}(x,y)$ by $O(\alpha)$. The asymptotics for $n\sim1/\alpha$ was obtained in \cite{Zh92,SunShen} (up to the change $\alpha$ to $\varepsilon$ as noted above) and has  the following form: in a small neighborhood of the point $\xi=0$ it is expressed through the Bessel functions  and $\varphi_n=O(1)$; in a small neighborhood of the turning point $\xi=q_{0}(\lambda)$ it is expressed through the Airy function and $\varphi_n=O(\alpha^{1/3})$; and in the inner points of the interval $\xi\in (0,q_{0}(\lambda))$ it is expressed through the standard WKB (or ray method) approximation and $\varphi_n=O(\alpha^{1/2})$. We will not perform the corresponding calculations because they are described  in \cite{Zh92,SunShen} and because these results can be easily obtained from (\ref{phi(x)}) by using the stationary phase method and its generalizations.

 Thus we restrict ourselves to the case $n=O(1)$ (cf. \cite{Zh92}). We will show that for such $n$ the asymptotics of (\ref{UB(p)}) is expressed in terms of the Laguerre polynomials. 
In this case the turning point $q_{0}(\lambda)$ tends to zero as $\lambda\to 0$.
Obviously, by (\ref{lam_{n}^2}), $\lambda=(2n+1)\alpha+O(\alpha^3)$ for $n=O(1)$.
We have, by (\ref{B(p,y)}), (\ref{Fi}), (\ref{q})   	 
\begin{align}\label{3-1-1}
& \Phi(p)=(2n+1)\alpha \arctan p+O(\alpha^2)
\\\nonumber\\\label{3-1-2}
&  b(p,y,\lambda)=\big(1+p^2\big)^{-1/2}+O(y).
\end{align}
Since $|e^{-ip\xi/\alpha}|\leq e^{-\abs{\min\limits_{c} \Im p}\xi/\alpha}$ and $0<y<\xi\alpha^{-1} \tan\alpha$, the $O(y)$ term in (\ref{3-1-2}) contributes an $O(\alpha)$ correction to the asymptotics of (\ref{UB(p)}) and we have 
\begin{equation*}
\begin{array}{lll}
\varphi_{n}(\xi,y)&=&-\ds\frac{1}{2\pi}\ds\oint\limits_{\mathcal{C}} e^{{-ip\xi}/{\alpha}+\frac{1}{2}(2n+1)\ln\frac{i-p}{i+p}}\;\ds\frac{1}{(p-i)^{1/2}(p+i)^{1/2}}\;dp+O(\alpha)\\\\
&=&\ds\frac{(-1)^{n-\frac{1}{2}}}{2\pi}\ds\oint\limits_{\mathcal{C}} e^{{-ip\xi}/{\alpha}}\;\ds\frac{(p-i)^{n}}{(p+i)^{n+1}}\;dp+O(\alpha).
\end{array}
\end{equation*}
Making the change of the variable $p=t-i$, we obtain
\begin{equation*}
\varphi_{n}(\xi,y)=\ds\frac{(-1)^{n-\frac{1}{2}}}{2\pi}\ds\int\limits_{\mathcal{C}+i} e^{{-i(t-i)\xi}/{\alpha}}\;\ds\frac{(t-2i)^{n}}{t^{n+1}}\;dt+O(\alpha).
\end{equation*}
Calculating the residue at the point $t=0$, we come to
\begin{equation*}
\varphi_{n}(x,y)=e^{-\frac{\xi}{\alpha}}(-1)^{n} \ds\frac{1}{n!}\Bigg(\ds\frac{d^{n}}{dt^{n}} e^{-it\xi/\alpha}(t-2i)^{n}\Bigg)\Bigg\vert_{t=0}+O(\alpha).
\end{equation*}
We have, denoting $t-2i=s$
\begin{equation*}
\begin{array}{lll}
\varphi_{n}(x,y)&=&e^{-\frac{\xi}{\alpha}}(-1)^{n} \Bigg(\ds\frac{1}{n!}\ds\frac{d^{n}}{ds^{n}} e^{-i(s+2i)\xi/\alpha}s^{n}\Bigg)\Bigg\vert_{s=-2i}+O(\alpha)\\\\
&=& (-1)^{n}\; e^{\xi/\alpha} \Bigg(\ds\frac{1}{n!}\ds\frac{d^{n}}{ds^{n}} e^{-is\xi/\alpha} s^{n}\Bigg)\Bigg\vert_{s=-2i}+O(\alpha).
\end{array}
\end{equation*}
Denoting $is\xi/\alpha=u$, we obtain 
\begin{equation}\label{Lag}
\varphi_{n}(x,y)= (-1)^{n}\; e^{-\xi/\alpha} \Bigg(e^{u}\;\ds\frac{1}{n!}\ds\frac{d^{n}}{du^{n}} e^{-u}\;u^{n}\Bigg)\Bigg\vert_{u=2\xi/\alpha}+O(\alpha).
\end{equation}
Hence finally
\begin{equation}
\varphi(x,y)=(-1)^{n}\; e^{-\xi/\alpha}\; {\rm L}_{n}\big(2\xi/\alpha\big)+O(\alpha),
\end{equation}
where $\xi=\alpha x$ and ${\rm L}_{n}(u)$ is the Laguerre polynomial defined by
\begin{equation*}
{\rm L}_{n}(u)= e^{u}\;\ds\frac{1}{n!} \ds\frac{d^{n}}{d u^{n}} e^{-u} u^{n}.
\end{equation*}
Note that the result (\ref{Lag}) coincides,  in the case of the straight bottom, with the formal asymptotics obtained in \cite{Miles}.

\section{Appendix\;1. Integral representation of the Ursell modes}

Here we prove (\ref{n>0}). It is obvious that the integrand in (\ref{var_{x,0}}) decreases rapidly for $0<\theta<\alpha$, $\rho>0$ and is meromorphic inside the contour $\Gamma$. Thus we can obtain (\ref{n>0}) using the residue theorem. The poles of the function $P_{n}$ lying between $\Gamma_{-\alpha}$ and $\Gamma_{-\pi}$ are   (see (\ref{w_k})
\begin{equation*}
w_{k}=-i\ds\frac{\pi}{2}+i\alpha(2n-2k-1),\quad k=0,1,\cdots,2k
\end{equation*}
and
\begin{equation}\label{r}
\overset{}{\underset{w=w_{k}}{\rm{Res}}}\;e^{-i\rho\sinh(w+i\theta)}\;P_{n}(w)= e^{-x\cos\big(\alpha(2n-2k-1)\big)+y\sin\big(\alpha(2n-2k-1)\big)}\cdot 2i\prod_{j=0,j\pm k}^{2n}\cot\big(\alpha(j-k)\big).
\end{equation}
Hence, representing $\varphi_{n}$ as the sum of three terms corresponding to $k=n$, $k=1,2,\cdots,n-1$ and $k=n+1, n+2,\cdots, 2k$, we obtain
\begin{equation}\label{d}
\varphi_{n}(x,y)=U_{n}^{(0)}(x,y)+U_{n}^{(1)}(x,y)+U_{n}^{(2)}(x,y),
\end{equation}
where
\begin{align*}
& U_{n}^{(0)}(x,y)=e^{-x\cos\alpha-y\sin\alpha}\;(-1)^{n}\prod_{j=0}^{n}\cot^{2}(\alpha j);\\\\
& U_{n}^{(1)}(x,y)=\sum_{k=0, j\neq k}^{n-1} e^{-x\cos\alpha(2n-2k-1)+y\sin\alpha(2n-2k-1)} \prod_{j=0, j\neq k}^{2n}\cot\Big(\alpha(j-k)\Big);\\\\
& U_{n}^{(2)}(x,y)=\sum_{k=n+1}^{2n} e^{-x\cos\alpha(2n-2k-1)+y\sin\alpha(2n-2k-1)} \prod_{j=0, j\neq k}^{2n}\cot\Big(\alpha(j-k)\Big).
\end{align*}	
Making the change of the index $k$
\begin{equation*}
m=n-k,~~ m=1,2, \cdots,n~~{\rm in}~~U_{n}^{(1)} \quad {\rm and}\quad m=-1,-2n,\cdots,-n~~{\rm in}~~U_{n}^{(2)},
\end{equation*}
we obtain
\begin{align*}
& U_{n}^{(1)}=\sum_{m=1}^{n} e^{-x\cos\alpha(2m-1)+y\sin\alpha(2m-1)} \prod_{j=0, j\neq n-m}^{2n}\cot\Big(\alpha(j-n+m)\Big)\\\\
& U_{n}^{(2)}=\sum_{m=1}^{n} e^{-x\cos\alpha(2m+1)-y\sin\alpha(2m+1)} \prod_{j=0, j\neq n-m}^{2n}\cot\Big(\alpha(j-n+m)\Big).
\end{align*}
Let us multiply  identity (\ref{d}) by $(-1)^{n}\tan^{2}(j\alpha)$. We obtain
\begin{equation*}
\begin{array}{lll}
(-1)^{n}\tan^{2}(j\alpha)\;\varphi_{n}(x,y)&=&e^{-x\cos\alpha-y\sin\alpha}+\ds\sum_{m=1}^{n}\;B_{mn}\Big(e^{-x\cos\alpha(2m-1)+y\sin\alpha(2m-1)}\Big)+\\\\
&+&\ds\sum_{m=1}^{n}\;B_{mn}\Big(e^{-x\cos\alpha(2m+1)-y\sin\alpha(2m+1)}\Big),
\end{array}
\end{equation*}
where
$$
B_{mn}=(-1)^{n}\tan^2(j\alpha)\prod_{j=0, j\neq n-m}^{2n}\cot\Big(\alpha(j-n+m)\Big),\quad m=1,2,\cdots, n. 
$$
It is easy to see that
 $B_{mn}=A_{mn}$,~$m=1,2,\cdots,n$, where $A_{mn}$ are given in (\ref{A_mn}). Hence (\ref{n>0}) follows.~~~~~~$\blacksquare$

\section{Appendix\;2. WKB approximation}

Here we briefly recall the construction of the WKB solutions to the problem (\ref{delV})-(\ref{lam^2}) \cite{Zh92, SunShen} and show that in the leading term they coincide with (\ref{phi(x)}). In the coordinates $\xi=\alpha x, y$ problem  (\ref{delV})-(\ref{lam^2}) reads 

\begin{subequations}
\begin{empheq}[left=\empheqlbrace]{align}
\alpha^2\varphi_{\xi\xi}+\varphi_{yy}-\varphi &= 0, \qquad\quad~ 0<y<\beta\xi, 
\label{A1}
\\ \nonumber\\
\varphi_{y}+\lambda\varphi &=0 , \quad\quad\quad~ y=0,
\label{A2}
\\ \nonumber\\
\varphi_{\xi}\;\alpha\tan\alpha-\varphi_{y} &= 0,   \qquad\quad ~y=\beta\xi
\label{A3} 
\end{empheq}
\end{subequations}
where $\beta=\alpha^{-1}\tan\alpha$. Following \cite{Miles}, we look for the solution in the form 
\begin{equation}\label{A4}
\varphi=\ds\frac{1}{2\pi}\ds\int\limits_{\mathcal{C}} e^{-ip\xi/\alpha}\Big(\cosh\tau y-\lambda\tau^{-1}\sinh\tau y\Big) f(p)dp,
\end{equation}
where $\tau=\sqrt{1+p^2}$, $f(p)$ is a new unknown, and $\mathcal{C}$ is a contour on the complex plane $p\in\C$. The expression (\ref{A4}) satisfies (\ref{A1}) and (\ref{A2}) exactly, and condition (\ref{A3}) gives the following integral equation for the function $f(p)$:    
\begin{equation}\label{A5}
\ds\frac{1}{2\pi}\ds\int\limits_{\mathcal{C}} e^{-ip\xi/\alpha}\; L(\xi,p) f(p) dp=0,
\end{equation}
where $L(\xi,p)=L_0(\xi,p)+i\alpha L_1(\xi,p)$,
\begin{equation}\label{L12}
L_{0}=\tau\sinh \tau\beta\xi-\lambda\cosh \tau\beta\xi,\quad L_1=\beta p\big(\cosh \tau\beta\xi-\lambda\tau^{-1}\sinh \tau\beta\xi\big).
\end{equation}
Since $\beta=1+O(\alpha^2)$ and we will construct only the leading term of the WKB asymptotics, we will assume that $\beta=1$ in (\ref{L12}). We look for the solution of (\ref{A5}) in the standard WKB-form
\begin{equation}\label{A6}
f(p)=a(p) e^{i\Phi(p)/\alpha}
\end{equation}
with $a(p)$ regular in $\alpha$ and $\Phi(p)$ analytic on $\C$. The stationary points of the integrand in (\ref{A5}) satisfy
\begin{equation*}
-\xi+\Phi'(p)=0;
\end{equation*}
thus, in the leading term, (\ref{A5}) vanishes if
\begin{equation}\label{A7}
L_{0}(\Phi',p)=0.
\end{equation}
 Equation (\ref{A7}) (the Hamilton-Jacobi equation) is equivalent to $H(\Phi',p)=\tau$, where the Hamiltonian $H(\xi,p)$ is given by $$H(\xi,p)=\tau\tanh \tau\xi$$, and hence $\Phi$ coincides with (\ref{Fi}) and, obviously, also depends on $\lambda$. Further,
\begin{equation}\label{A8}
\begin{array}{lll}
\ds\int\limits_{\mathcal{C}}e^{-ip\xi/\alpha+i\Phi/\alpha}\Big(L_0(\xi,p)+i\alpha L_1(\xi,p)\Big) a(p)\;dp\\\\
=\ds\int\limits_{\mathcal{C}} e^{-ip\xi/\alpha+i\Phi/\alpha}\Big(L_{0}(\xi,p)-L_{0}(\Phi',p)\Big) a(p)\;dp+i\alpha\ds\int\limits_{\mathcal{C}} e^{-ip\xi/\alpha+i\Phi/\alpha}\;L_{1}(\tau,p)a(p)\;dp.
\end{array}
\end{equation}
Integrating by parts in the first integral using the formula 
\begin{equation*}
\ds\frac{\alpha}{-i\xi+i\Phi'}\;\ds\frac{\partial}{\partial p}e^{-ip\xi/\alpha+i\Phi/\alpha}=e^{-ip\xi/\alpha+i\Phi/\alpha}
\end{equation*}
we obtain that (\ref{A5}) reduces to
\begin{equation*}
\ds\frac{i\alpha}{2\pi}\ds\int\limits_{\mathcal{C}} e^{-ip\xi/\alpha+i\Phi/\alpha}\Bigg\lbrace-\ds\frac{\partial}{\partial p}\Bigg(\ds\frac{L_0(\xi,p)-L_0(\Phi',p)}{\xi-\Phi'} a
(p)\Bigg)+L_1(\xi,p) a(p)\Bigg\rbrace\;dp
\end{equation*}
and this expression vanishes in the leading term if
\begin{equation*}
-\ds\frac{\partial}{\partial p}\Bigg(\ds\frac{L_0(\xi,p)-L_0(\Phi',p)}{\xi-\Phi'} a(p)\Bigg)+L_1(\xi,p) b(p)=0~~{\rm at}~~\xi=\Phi'.
\end{equation*}
This gives an equation for $a(p)$ (the so-called transport equation):
\begin{equation*}
-L_{0\xi}\; a'-L_{0\xi p}\; a-\ds\frac{1}{2}L_{0\xi\xi}\; \Phi'' a+L_1 a=0,\quad\xi=\Phi'.
\end{equation*}
Expressing the derivatives of $L_0$ in terms of the derivatives of $H$, after some algebraic manipulations, one comes to 
\begin{equation*}
-\cosh\tau \Phi'\sqrt{H_{\xi}\big(\Phi',p\big)}\;\ds\frac{d}{dp}\sqrt{H_{\xi}\big(\Phi',p\big)}\;a(p)=0,
\end{equation*}
and hence
\begin{equation}\label{A9}
a(p)=\ds\frac{1}{\sqrt{H_{\xi}(\Phi',p)}}=\ds\frac{\cosh\tau\Phi'}{\tau}=\ds\frac{1}{\tau}\;\ds\frac{1}{\big(1-\lambda^2/\tau^2\big)^{1/2}}
\end{equation}
and, substituting in (\ref{A4}), we see that (\ref{A4}) coincides with (\ref{phi(x)}).~~~~$\blacksquare$

\medskip
The edge-wave condition $\abs{\varphi(0,0)}<\infty$ requires that the ``quantization condition'' $\ds\frac{1}{\alpha}\ds\int\limits_{-\infty}^{\infty} \Phi'(p)\;dp=(2n+1)\pi$ be valid, this means that $\lambda$ satisfies (\ref{lam_{n}^2}) so that the contour $\mathcal{C}$ in (\ref{A4}) can be deformed into a closed contour shown in Fig.\;\ref{C1} (see Section\;3.1).

\section{Conflict of interest}

The authors have no competing interests to declare that are relevant to the content of this article.

\section{Data availability statement}

No datasets were generated or analysed during the current study.

\section{Funding statement}

PZ and AM  were partially financially supported by Sistema Nacional de Investigadores (grants 14536 and 21641), and  MIRR was partially financially supported by Vicerrector\'ia de la Investigaci\'on de la Universidad Militar Nueva Granada (grant IMP-CIAS 4084,  2025/1-2026/2).

\section{Acknowledgement}
The authors are grateful to Dr. Jos\'e Eligio De la Paz M\'endez for technical assistance.

\end{document}